\documentclass[12pt,a4paper]{article}
\usepackage[utf8]{inputenc}
\usepackage[english]{babel}
\usepackage{amsmath}
\usepackage{amsfonts}
\usepackage{amssymb}
\usepackage{makeidx}
\usepackage{graphicx}
\usepackage{morefloats}
\usepackage[left=2cm,right=2cm,top=2cm,bottom=2cm]{geometry}
\usepackage{pdflscape}

\author{Arindam Ghosh$^{1,*}$, Indrani Banerjee$^{2}$, and Sandip K. Chakrabarti$^{3}$}
\title{Does Cyg X-1 have a Small Accretion Disc?}

\begin{document}

\maketitle
\begin{center}
{$^1$S. N. Bose National Centre for Basic Sciences, Salt Lake, Kolkata 700106\\ 
$^{2}$Indian Association for the Cultivation of Science, Raja S. C. Mullick Road, Kolkata 700032\\
$^3$Indian Centre for Space Physics, Chalantika 43, Garia Station Rd., Kolkata 700084\\
India\\
{\it arindam.ghosh@bose.res.in; tpib@iacs.res.in; sandip@csp.res.in}\\}
\end{center}



\abstract{We analyze several outbursts of a few transient sources using Proportional Counter Array (PCA) data (2.5-25 keV) as well as All Sky Monitor (ASM) data (1.5-12 keV) of Rossi X-ray Timing Explorer (RXTE) satellite. We find a time delay between the arrival times of the Keplerian disc component and the halo of the Two-Component Advective Flow (TCAF) when the spectral data is fitted with TCAF solution. We compare this time delay from the spectral fits with the TCAF solution of the transient 
low mass X-ray binaries (LMXBs) e.g., GX 339-4, H 1743-322 and MAXI J1836-194 with that of the high mass X-ray Binary (HMXB), Cyg X-1. We find that several days of time delays are observed in LMXBs while for Cyg X-1 the delay is negligible. 
We interpret the large delay to be due to the viscous delay of a large Keplerian component to reach the inner region as compared to 
nearly free-fall time taken by the low angular momentum halo component. 
The delay is of the order of a few days for the low mass X-ray binaries (LMXBs) where the feeding 
is primarily through the Roche-lobe. However, it is negligible in a wind-fed system like Cyg X-1 since a very small Keplerian disc is created here by slowly redistributing the low angular momentum of the wind. As a consequence, sporadic soft or intermediate spectral states are observed.}




\section{Introduction} 

Accretion process around stellar mass black holes is an important subject since the observations in all the energy bands could be interpreted well if we have a clear understanding of how matter actually leaves the companion and plunges into the primary compact object. The well known work of Shakura \& Sunyaev (1973) proposes that matter leaving the companion through the Roche lobe forms a Keplerian disc till the last stable circular orbit located at, say, $3r_g$, where $r_g=2GM_{\bullet}/c^2$, the Schwarzschild radius for a non-rotating black hole of mass $M_{\bullet}$. This component is responsible for producing a multi-colour black body spectrum. However, with the advent of satellite observations it has become clear that there is another component called the Compton cloud, which is made up of hot electrons, that is responsible for producing the power-law component of the spectrum, especially in hard states, i.e., the states dominated by high energy photons. The literature has numerous models and locations for this cloud, e.g. a magnetic corona (Galeev et al. 1979), a hot gas corona over the disc (Haardt \& Maraschi 1993; Narayan \& Yi 1994; etc.), a plasma region partially covering the disc (Zdziarski et al. 2003). 
However, the biggest problem was how to regenerate and sustain the energy of the Compton cloud as and when observed. Chakrabarti \& Titarchuk (1995) proposed that instead of a static corona, one could have a dynamic corona if the accretion flow has two-components 
such that the low angular momentum sub-Keplerian flow surrounds the Keplerian disc that lies in the  equatorial plane. The advective sub-Keplerian component, also known as the sub-Keplerian halo, falls almost radially and soon becomes supersonic. However, at some distance from the black hole the centrifugal pressure dominates over gravity slowing down the flow such that its velocity becomes subsonic through a centrifugal pressure dominated shock transition (Chakrabarti 1989; Chakrabarti 1996). 
The loss of kinetic energy is transformed to heat which distends the flow vertically resulting in the formation of a shock which is the 
outer edge of sub-sonic region called the CENtrifugal barrier supported BOundary Layer (CENBOL). This disaggregation of two components 
is possible when there is a vertical viscosity parameter gradient. 
In this two component advective flow (TCAF) paradigm, the CENBOL behaves as the Compton cloud, where the electrons scatter the intercepted soft photons from the Keplerian disc and inverse-Comptonizes them to produce a power-law component of the spectrum following a process described in Sunyaev \& Titarchuk (1980, 1985).\\

The standard disc in the equatorial plane has a Keplerian distribution which, if extended all the way from the outer edge, would mean 
that matter must remove more than 99\% of its angular momentum in order to fall into the black hole with $l\sim l_{ms}$, 
the angular momentum at the marginally stable value. This requires a very high viscosity. On the other hand, if low angular momentum matter
is injected from the outer edge or from a `pile-up' radius (the radius till which Keplerian flow could manage to keep the 
Shakura-Sunyaev disc configuration from outer edge) it will flow till the horizon and form the CENBOL in presence of a weak viscosity.
However, in presence of a super-critical viscosity on the equatorial plane,
a Keplerian disc could be produced in between the pile-up radius and the CENBOL boundary (shock). 
The changes in solution topology below and above a critical value with a Shakura-Sunyaev (1973) viscosity parameter $\alpha$ have been studied extensively (see, Chakrabarti 1990; Chakrabarti 1996). 
The formation of TCAF through numerical simulations is described in details in Giri \& Chakrabarti (2013). 

A TCAF solution requires four independent physical parameters to fit a spectrum. They are the two accretion rates (of halo and disc), the shock location and its strength $R$ (compression ratio). The shock location gives the size of the CENBOL or the Compton cloud. Together with $R$ and accretion rates, we get the optical depth. The soft photons are supplied by the disc while the CENBOL, enriched with constantly replenished hot electrons by the halo, emits the power-law component of radiation (hard photons). Inner boundary of the truncated disc coincides with the shock location. Thus, TCAF recycles a few parameters to do various tasks while giving the complete description of the spectrum. Moreover, since the electron number 
density and photon densities require a knowedge of the mass of the blackhole, TCAF also can estimate the mass, if it is not known a priori.
RXTE data of several years has been used to show that one requires two components to explain time lags or delays for high mass and low mass X-ray binaries. Smith et al. (2001, 2002) came to this conclusion, purely from the spectral data analysis, that the time lag between an empirical power-law index (which was extracted from RXTE/PCA data and was shown to be proportional to the inverse of hardness ratio $\mathrm{HR\mbox2}^{-1}$) and hard photon flux (RXTE/ASM $3-12$ keV or RXTE/PCA $2.5-25$ keV) is very low (about a day) in HMXBs, but of the order of a few weeks in LMXBs. This low time lag in HMXBs was assumed to be due to a smaller Keplerian disc.\\

Recently, in Ghosh and Chakrabarti (2018), a similar results are obtained directly from a temporal analysis with long time ($\sim13$ years) RXTE/ASM data ($1.5-12$ keV) of Cyg X-1 (including its several {\it flare}s), and the data of several outbursts in transient LMXBs like, GX 339-4, XTE J1650-500, etc. A generic power-law photon index, which also traces the spectral evolution of all outbursts as a dynamical indicator, was used there. Keplerian matter from the companion continues to pile up at some distance (the pile-up or accumulation radius, $X_{a}$) from the black hole until there is a dynamic rise of viscosity. About ten outbursts were considered to show the arrival time delays (ranging from $0$ day to a few weeks) of the disc flow. These delays are proportional to the accumulation radii ($X_{a}$s) if one assumes the viscosity parameter remains the same. They are also a measure of energy release and duration of the outburst (the declining phase, in particular). However, for Cyg X-1, it has been concluded (Ghosh and Chakrabarti 2018) that such a time lag is $\sim 0$ day and could be less than $\sim5$ hours. 
Earlier, a totally different approach was taken in Ghosh and Chakrabarti (2016) to investigate the size of the disc. There they took the Fourier spectra (power density spectra and periodograms) of total ($1.5-12$ keV) RXTE/ASM as well as Swift/BAT ($15-50$ keV) lightcurves and showed that sharp peaks arose at the orbital periods in the case of HMXBs (e.g., Cyg X-1) while very broad peaks close to the orbital periods ensued for LMXBs. This was interpreted to be due to viscous smearing of mass accretion rate variations at the orbital period, possibly because of an eccentricity in the binary orbit. The smaller disc size spares the dominant wind accretion in HMXBs from such an effect and allows the carriage of orbital information to the inner edge. In LMXBs, the longer travel time through a larger disc leads to smearing out of the aforesaid variations.\\ 

In the present paper, we take yet another approach, this time using the RXTE/PCA data to show that when the spectral data is fitted with TCAF and two accretion rates are extracted from the fits, their time evolution clearly indicates that they peak at different times - always the low angular momentum flow rate peaks before the Keplerian disc rate in the rising phase of an outburst. This we interpret to be due to viscous travel time 
in the Keplerian component since the halo component takes negligible time. By examining the published results of TCAF fits for a few LMXBs viz. MAXI J1836-194 (Jana, Debnath, Chakrabarti, Mondal, \& Molla 2016), GX 339-4 (Debnath, Mondal, and Chakrabarti 2015), H 1743-322 (Molla, Chakrabarti, Debnath \& Mondal 2017) and as well as our analysis of Cyg X-1 with TCAF, we clearly show that the arrival time delay between the two components is negligibly small in the latter, while it is of about a week in the former three LMXBs. This indicates that Cyg X-1 could harbour a very small Keplerian disc, the size of which varies over spectral states. In harder states, our TCAF fits clearly show various inner truncation radii. 

In the next Section, we present the procedure of analysis of the RXTE/ASM and RXTE/PCA data using TCAF as the additive model.  
In \S3, we present a comparison of the time evolution of mass accretion rates in LMXBs using published results 
and the way these quantities vary in Cyg X-1. Finally in \S4, we summarize our results. 

\section{Data Analysis}

We examine four outbursts in three transient LMXBs, viz., MAXI J1836-194, GX 339-4, H 1743-322, and with both RXTE/PCA (2.5-25 keV) data and RXTE/ASM (1.5-12 keV) data. Besides, two outburst-like `flare's in the HMXB Cyg X-1 are examined in greater detail. 

Fitting of spectral data using TCAF as an additive model has been carried out by Jana et al. (2016), Debnath, Mondal, \& Chakrabarti (2015), and Molla, et al. (2017) for the outbursts in MAXI J1836-194, GX 339-4, and H 1743-322 respectively. We do not repeat the analysis here but invoke their results to compare with our results obtained in this paper. Each fit of a spectrum, apart from the mass of the black hole itself,  enables us to extract four independent parameters, 
which are (i) the standard Keplerian disc accretion rate ($\dot{m}_{disc}$), (ii) the low angular momentum halo 
acretion rate ($\dot{m}_{halo}$), (iii) the location of the Centrifugal barrier ($X_s$) (which also represents the 
size of the Compton cloud or CENBOL), and finally, (iv) the shock compression ratio ($R$), 
which indirectly gives the opacity of the Compton cloud. While the aforesaid four quantities generate 
the shape of the overall spectrum, suitable model normalization $N$ is used to raise or lower the whole spectra 
in order to reproduce the observed spectrum. Since $N(=\frac{{r_g}^2}{4\pi D^2}\sin i)$ depends on the mass of the black hole, the distance $D$ (in units of 10 kpc) of the source and the inclination angle $i$,  it is expected to remain roughly constant across the 
spectral states. This is because TCAF generates the spectrum in its entirety which includes hard, soft and reflected components.
Thus a single constant factor $N$ is needed to fit any spectra for a given object and instrument pair. 
With spectral evolution, we obtain the time evolution of $\dot{m}_{disc}$, $\dot{m}_{halo}$, $X_s$ and $R$. 
After triggering of an outburst, matter rushes in, and both the rates start to rise. 
Our goal is to find the times at which these two mass accretion rates achieve maximum 
values during the rising state. If these two times are near simultaneous, it means that the standard disc is small in size. 

For Cyg X-1, we analyze the RXTE/PCA archival data of about 18 months (MJD 52621 to MJD 53151) which includes two of its flaring phases. HEASARC's software packages HEASOFT (version HEADAS 6.18) and XSPEC (version 12.9.0) are used for the purpose. Data collected for elevation angles greater than $10^o$, for offset less than $0.02^o$, and those acquired during the South Atlantic Anomaly (SAA) passage, are excluded. {\it Standard2} mode Science Data of PCA (FS4a*.gz) with $16s$ spectral binning are used. The {\sc runpcabackest} task is run to extract the PCA background spectra by using the latest bright-source background model. PCA breakdown correction and dead-time correction are done. The response files are prepared using the task {\sc pcarsp}. Systematic error of $1\%$ is included. For each data ID, we have extracted the spectra from PCU2 data;  ($2.5-25$ keV) PCA spectra are background subtracted and fitted with local additive model {\sc fits} files generated from TCAF solution as described in Debnath, Chakrabarti, \& Mondal (2014). In order to obtain the best fit, the interstellar absorption of the hydrogen column densities $N_H$ (see column 4 in Tables 1 \& 2) in the unit of $10^{22} cm^{-2}$ (Grinberg et al. 2015) is taken for the absorption model \textit{phabs}. The data is analyzed using the TCAF solution to obtain various parameters of the accretion flow. In order to fit the spectrum using a TCAF-based model, one needs to provide the four aforesaid input parameters. The accretion rates are in units of the Eddington rate ($\dot M_{Edd}=1.4\times 10^{17} M_{\bullet}~ gs^{-1} $), the location of the shock is in units of Schwarzschild radius ($r_g=2 G M_{\bullet}/c^2$) and the shock compression ratio, $R$, is dimensionless. The black hole mass $M_{\bullet}$ (or $M_{BH}$) can be obtained from our fit as well if not known before and it would be in units of the solar mass $M_{\odot}$. Apart from these, if the normalization factor for the object is unknown, we need to supply and adjust till it is nearly constant for all the observations of the same object while keeping the reduced $\chi^2$ to a minimum. 

Archival RXTE/ASM daily-average lightcurve data, wherever available, are used to compare the rates of photon fluxes with the mass accretion rates obtained with RXTE/PCA data ($2.5-25$ keV). The RXTE/ASM operates over $1.5-12$ keV energy range. It comprises of three energy bands, viz. A=($1.5-3$ keV), B=($3-5$ keV), \& C=($5-12$ keV) respectively. If $a$, $b$, \& $c$ are the number of photons in A, B, \& C bands respectively, then the A-band represents low-energy photons $a$ (or soft flux), B \& C bands together represent the Comptonized, high-energy photon counts (or, hard flux) of $(b+c)$. Along with these photon fluxes two hardness ratios, defined here as, $\mathrm{HR\mbox1}=(b/a)$, $\mathrm{HR\mbox2}=(c/b)$, and the Comptonizing efficiency (Pal \& Chakrabarti 2015) defined as CE=$(b+c)/a$ (Ghosh \& Chakrabarti 2018) are also examined. 

Accretion mass rates are also cross-correlated wherever reasonable in order find the desired time lag between them until the peak outburst/flare is achieved. \\

\section{Results}

The Keplerian and the sub-Keplerian accretion rates during an outburst in MAXI J1836-194 are obtained by Jana et al. (2016). These are plotted in Fig. 1. Accretion rate ratio (ARR=$\dot{m}_{halo}/\dot{m}_{disc}$) is also drawn to show the relative importance of one rate with respect to the other. The horizontal axis shows time in days. Though both the rates are seen to increase with time, the halo rate peaks about $9d$ before the disc rate. On the day the disc rate is maximum, the halo rate is found to be minimum, clearly indicating conversion of halo component to the disc component and vice versa depending on viscosity. Figure 2 shows an outburst in GX 339-4. Figure 2a shows the variation of hard and soft photon fluxes with RXTE/ASM daily average data. CE is also shown. Mass accretion rates, obtained with RXTE/PCA data by Debnath, Mondal, and Chakrabarti (2015), are plotted in Fig. 2b along with ARR in the rising phase. This outburst lasted for a long time and it stayed in the soft state for several months. However, from Figs. 2a-b, we see that the soft flux and the disc rate peak later as compared to the hard flux and the halo rate respectively. The time delays are respectively $6d$ and $7d$. In Figs. 3a-b, results of two outbursts of H 1743-322 are shown, where accretion rates and ARR are plotted. These results are taken from Molla et al. (2017). In both outbursts of H 1743-322, ARR reaches its maximum value of the order of $\sim 60$ (not shown for clarity) at their commencement or culmination. The shape of these rate variations are strong functions of viscosity in the flow and the exact shape would be difficult to predict. However, the common trend remains the same as in Fig. 1. Here, in Fig. 3a, the halo rate reaches the peak about $6d$ prior to peaking of the disc rate. In Fig. 3b, there are two overlapping outbursts, as we can see two sharp peaks in both the halo and disc rates and they are at about a week apart (marked).

So, all these outbursts of the LMXBs point to the same physical process that the viscous timescale inside the Keplerian disc is longer than that in the sub-Keplerian component by about  a week or so. This is clearly due to higher viscosity, that is prevalent in Keplerian discs required to transport the angular momentum efficiently.\\

We now turn our attention towards Cyg X-1. It normally stays in the so-called hard states, though it may visit the soft state quite irregularly and often for a very short time period. Relative increase in the count rates and fluxes are not as dramatic as in transient LMXBs, and because of their general irregularity we term them as `flare'(s) rather than outbursts. In Fig. 4, we present the long-time ($\sim9$ years, MJD 50500-53700) behaviour of Cyg X-1, in terms of (a) hard flux, (b) soft flux, (c) HR1, (d) HR2, and (e) CE. Weekly running-mean data of RXTE/ASM are used. Figures 4(a-e) illustrate that the soft flux slowly becomes comparable to the hard flux, and HRs \& CE gradually diminish in the course of the first 
$2000$ days, thereby indicating a softening of spectrum from hard to hard-/soft- intermediate state. Thereafter the spectrum becomes 
hard for about $100$ days, and a uniquely prominent outburst-like flare (marked I), with similar photon fluxes, occurs. In fact, throughout the duration of these 9 years, there is not a single flare which is as conspicuous as `Flare I' and this motivates us to analyze this flare in a greater detail to understand the underlying accretion flow dynamics. We also study an atypical `Flare II', the flare next to `Flare I', because these two flares together might give a better insight of the Cyg X-1 disc. Both flares are zoomed in Fig. 5a, where the soft (1.5-3 keV) and hard (3-12 keV) photon counts are plotted with RXTE/ASM daily average data.
In Fig. 5b, we plot CE, HR1 \& HR2, corresponding to the lightcurves in Fig. 5a. We then fit the 2.5-25 keV spectrum of Cyg X-1 obtained from RXTE/PCA data for $\sim$ 530 days (MJD 52621-53151), which includes Flare I and Flare II, with TCAF solution and extract the corresponding accretion flow parameters, namely the disc rate $\dot{m}_{disc}$, the halo rate $\dot{m}_{halo}$, the location $X_s$ and the strength $R$ of the shock. The TCAF fit parameters of our concern of Flare I and Flare II are plotted along with the error bars. The details of flow parameters are given at the end in Table 1 \& Table 2, respectively for Flare I \& Flare II, despite a common neighbourhood in between. Just before Flare I the object was in the hard state (MJD 52621-52649). Figure 5b clearly shows the higher values of hardness ratios, HR1 \& HR2, and CE during this period. One can also see this from Fig. 5a, when the hard flux is greater than the soft flux over this duration. 
The results from the TCAF fits also corroborate the same since $\dot{m}_{halo}$ is consistently higher than $\dot{m}_{disc}$ during this period, i.e., the average halo rate is $1.771$ and the average disc rate is $0.717$ $\dot{M}_{Edd}$ during this period. Thus ARR ($\dot{m}_{halo}/\dot{m}_{disc}$) is also high during this period. The average location of the shock is $\sim 72.725 R_g$ while the average shock strength is $\sim 1.204$. Then suddenly over a span of $\sim 1.5$ hours, on MJD 52649.4852, both $\dot{m}_{halo}$ and $\dot{m}_{disc}$ rise simultaneously. The arrival time delay between them is $0d^{+4.42}_{-1.43}h$. This agrees with the conclusions of Ghosh \& Chakrabarti (2018). For LMXBs, we have seen that the halo rate always rises earlier than the disc rate, since the sub-Keplerian matter in the halo falls in free-fall timescale and the Keplerian matter falls in viscous timescale, the Keplerian matter should take longer time to accrete. However, if the two timescales are almost coincident it implies that the Keplerian disc is indeed small in size. This is a unique feature of wind-fed Cyg X-1. \\ 

During MJD 52649.4852-52735.4317, the disc rate and the halo rate do not exhibit much changes with an average $\dot{m}_{disc} \sim 2.389 \dot{M}_{Edd}$ and $\dot{m}_{halo} \sim 6.337 \dot{M}_{Edd}$. HR1, HR2, and CE continue to remain high throughout this duration indicating a hard-intermediate state. 
On MJD 52748.2531, the disc rate exceeds the halo rate and ARR drops indicating a transition to the soft state. It is difficult to specify how quickly this transition occurs as there is no available data between MJD 52735.4317 and MJD 52748.2531. Once it attains the soft state it remains there till MJD 52875.0352. The soft state is further characterized by a weak shock with an average strength of $\sim 1.095$ and the shock location $X_s$ approaching close to the black hole (the average shock location being $\sim 33.394~R_g$). This is evident from Table I and Fig. 5c. The onset of the soft state is further corroborated from Fig. 5b which indicates a drop in HR1, HR2, and CE over this duration. Thereafter, from MJD 52889.7476 to MJD 52963.500 the intermediate/hard state sets in with average $\dot{m}_{disc} \sim 2.729 \dot{M}_{Edd}$ which is less than the average $\dot{m}_{halo} \sim 6.328 \dot{M}_{Edd}$. Thus ARR goes up and the shock also moves outward with enhanced strength. From Fig. 5b one can see that during this time HR1, HR2, and CE also go up.\\

MJD 52973.649 marks the onset of Flare II when the disc rate again exceeds the halo rate, the ARR drops and the shock strength decreases from $3.113$ to $1.097$ and the shock location moves inward from $196.622 ~R_g$ to $35.255 ~R_g$ in the rising phase. This behavior continues till MJD 53055.318. Fig. 5b shows a decreased CE, HR1 and HR2 during this time period which is characteristic of a soft state.
The transition from the soft state to the intermediate/hard state occurs around MJD 53073.268 and the flow continues to exhibit such a behavior till MJD 53150.344. The average halo rate during this period is $\sim 6.322 \dot{M}_{Edd}$, the average disc rate $\sim 2.613 \dot{M}_{Edd}$, the average location of the shock $X_s \sim 193.933 R_g$ while the shock strength $R\sim 2.968$. With the commencement of the intermediate/hard state HR1, HR2, CE and ARR go up, which are evident from Figs. 5b and 5c respectively.\\

In view of Fig. 5c, it seems that Flare I begins in the hard state, whereas Flare II occurs immediately after Flare I without passing through the hard state in between. Rather, Flare II possibly begins in the soft-/hard- intermediate state of the declining phase of Flare I. It appears as if the Keplerian disc of Flare I did not recede totally. Since the changes in the Keplerian rate and the recession of the inner edge of the disc is decided by 
the viscosity at the accumulation radius, it is possible that a fresh but ephemeral ring-like disc forms from the halo component due to a sudden rise in viscosity which gives rise to Flare II. Therefore, Flare II is different from Flare I, even though both ends up with similar values of mass rates (compare at $\sim 300d$ \& $\sim 500d$ with respect to MJD 52610 in Fig. 5c). The subsequent rise and fall in mass accretion rates are akin to soft intermediate to soft states or vice versa in outbursts (Figs. 1-3). Thus the rates vary similarly. However, there is practically no delay in the peaking of the two photon fluxes (compare Fig. 5a with Fig. 2a). Therefore, the Keplerian disc size in Cyg X-1 could be very small and the viscous timescale through the Keplerian disc does not significantly enhance the matter travel time from the outer edge to the inner edge of the disc. This analysis further reinforces our conclusion that the size of Cyg X-1 disc is determined by totally different considerations from that of the transient LMXB sources we considered.
 
When the two mass accretion rates are cross-correlated, keeping the time window confined between the commencement and the peak of the flare/outburst, we also obtain the time lags similar to those estimated earlier. These are shown in Fig. 6. 

Although the mass estimation of the black hole in Cyg X-1 is not the goal of the present paper it is obtained as a `by-product'. As computed from both flares of Cyg X-1, the mass can be constrained well within the range $(14-15)M_{\odot}$, as is evident from the $7^{th}$  column in Table 1 \& Table 2. Our estimate is consistent with a recent finding of $14.81 \pm 0.98~ M_{\odot}$ from a dynamical study (Orosz et al. 2011). It is important to note here that using the supposed universal scaling between the intensity and the QPO frequency, Shaposhnikov \& Titarchuk (2007) earlier obtained a mass of $8.7 \pm 0.8~ M_{\odot}$, which is much lower. It is unclear at this stage if the crucial turnover QPO frequency in the latter model was determined accurately. Furthermore, their estimation of mass depends on the mass of another black hole as an input parameter and so on. However, this is beyond the scope of the present analysis.

\section{Summary}

In the present paper, based on TCAF model fit of the spectra during this rising phase of several outbursts, we have shown that the two accretion rates attain their peak values at two different times for LMXBs (viz. MAXI J1836-194, GX 339-4, and H 1743-322). The time gap between attainment of the two peaks for the LMXBs is found to be about a week or so. However, for HMXBs such as Cyg X-1, this time delay is negligible. This is confirmed by fitting the RXTE/PCA data 
of Cyg X-1 using TCAF solution where the accretion rates of the two components are found to peak on the same day. To our knowledge this is the first time that accretion rates of Cyg X-1 were directly extracted from the data across the spectral states and the conclusion about the disc size is made directly from the flow dynamics.

It is well known that the LMXBs are Roche lobe accretors. Matter coming from the companion may pile up at some intermediate radius due to a lack of viscosity and only after the viscosity is increased, possibly due to thermal-viscous instability as in novae outbursts, the flow rushes towards the black hole causing outbursts. Spectral analysis of the observed radiation shows that this flow could be having two components 
which reach peak values at two different times. In the case of HMXBs, such as Cyg X-1, we find that this gap is minimal. 
This only means that there is mostly the halo component, and the other disc component is very small in size. 
As Cyg X-1 is an wind accretor, the main component could only be the advective halo component of low angular momentum. 
This object does not show regular outbursts like LMXBs, but irregular `flares' (`mini-outbursts'), 
which could lead to persistent brightness of Cyg X-1. Hence it seems that there is no well-defined pile-up radius of the Keplerian matter.
The presence of a small Keplerian disc during `flares' of Cyg X-1 could only be interpreted as what is produced by some temporary 
increase in viscosity. 
Close to the black hole, the gas has high thermal pressure and even a small viscosity parameter could give rise to high 
viscous stress to transport angular momentum fast enough to generate a Keplerian distribution and create a small sized standard disc.
This has been shown by Giri \& Chakrabarti (2013) and Roy \& Chakrabarti (2017) using numerical simulations. 
We believe that what we find here for Cyg X-1 is a generic feature of any wind accretor. 
This will be investigated in future and reported elsewhere.
 
\section*{Acknowledgement}

The authors are thankful to NASA Archives for RXTE/ASM \& RXTE/PCA public data and facilities. They are also thankful to Dr. D. Debnath for providing the TCAF fits files,$TACF\_v0.3\_R4.fits$, which were used to fit Cyg X-1 data here.

\newpage

\begin{figure}
\begin{center}
\includegraphics[width=\columnwidth]{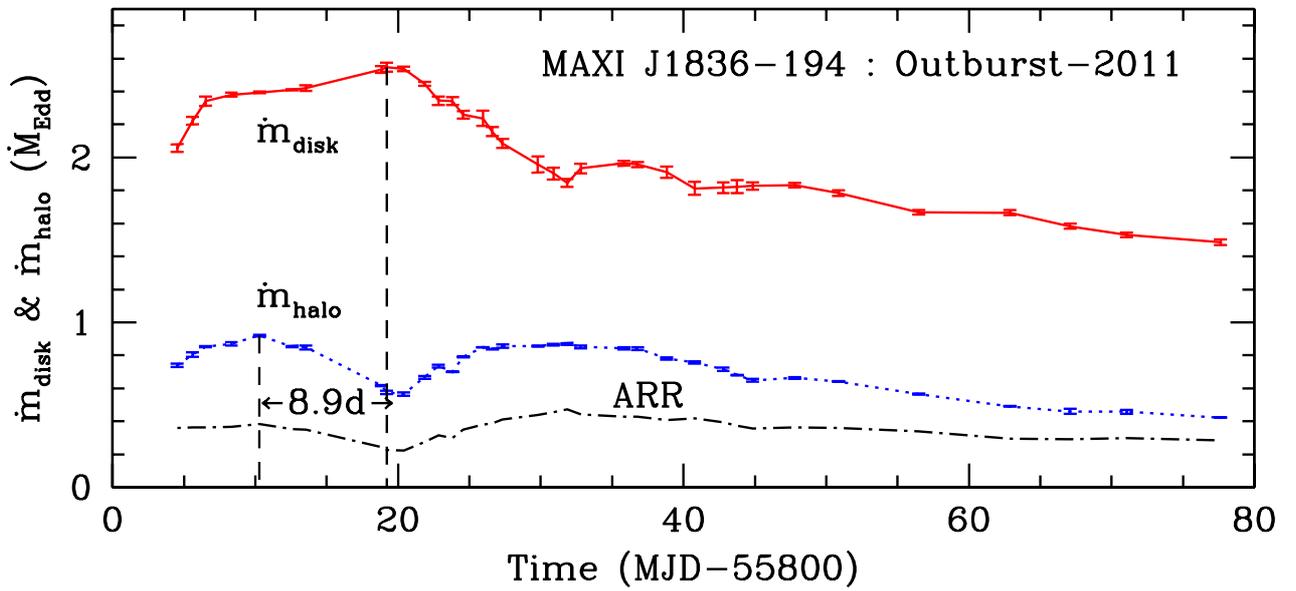}
\caption{Outburst in MAXI J1836-194: Mass accretion rates, computed with RXTE/PCA (2.5-25 keV) data, as given in Jana, Debnath, Chakrabarti, Mondal, \& Molla (2016), are plotted in (a). ARR represents the accretion rate ratio of $\dot m_{halo}$ to $\dot m_{disc}$. A time delay of $8.9d$ is marked.}
\end{center}
\end{figure}

\begin{figure}
\begin{center}
\includegraphics[width=\columnwidth]{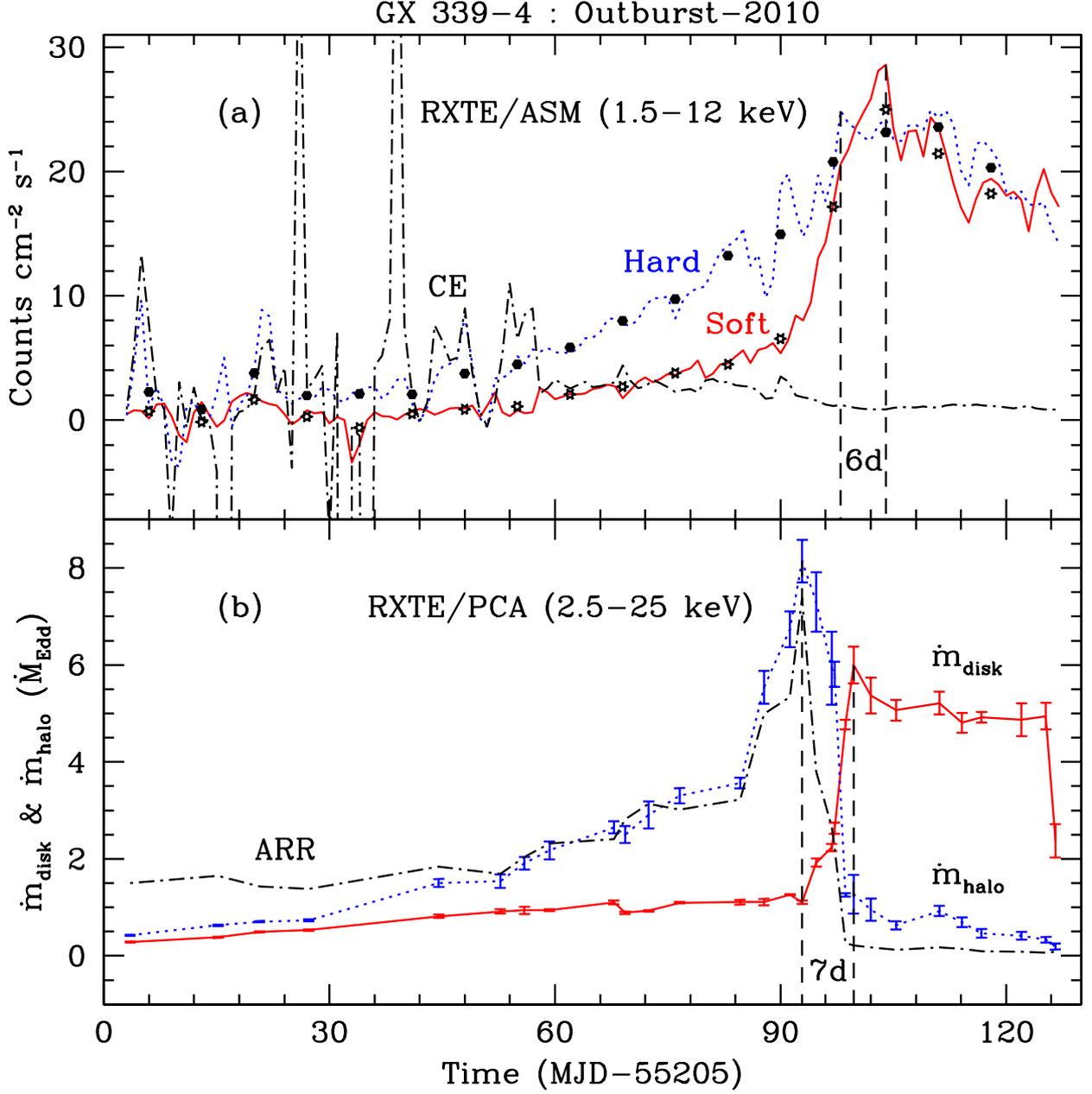}
\caption{Outburst in GX 339-4: Lightcurves of hard (3-12 keV) and soft (1.5-3 keV) photon fluxes from RXTE/ASM daily mean data are plotted in (a) in the rising phase. Weakly average data are also plotted with points (dots for hard \& stars for soft). Corresponding mass accretion rates, namely $\dot m_{disc}$ and $\dot m_{halo}$, computed with RXTE/PCA (2.5-25 keV) data by Debnath, Mondal, and Chakrabarti (2015), are plotted. Comptonization efficiency (CE) and the accretion rate ratio (ARR) are also drawn in (a) \& (b) respectively. Higher values of CE are ignored for clarity. Delay times of about a week are marked in each box.}
\end{center}
\end{figure}


\begin{figure}
\begin{center}
\includegraphics[width=\columnwidth]{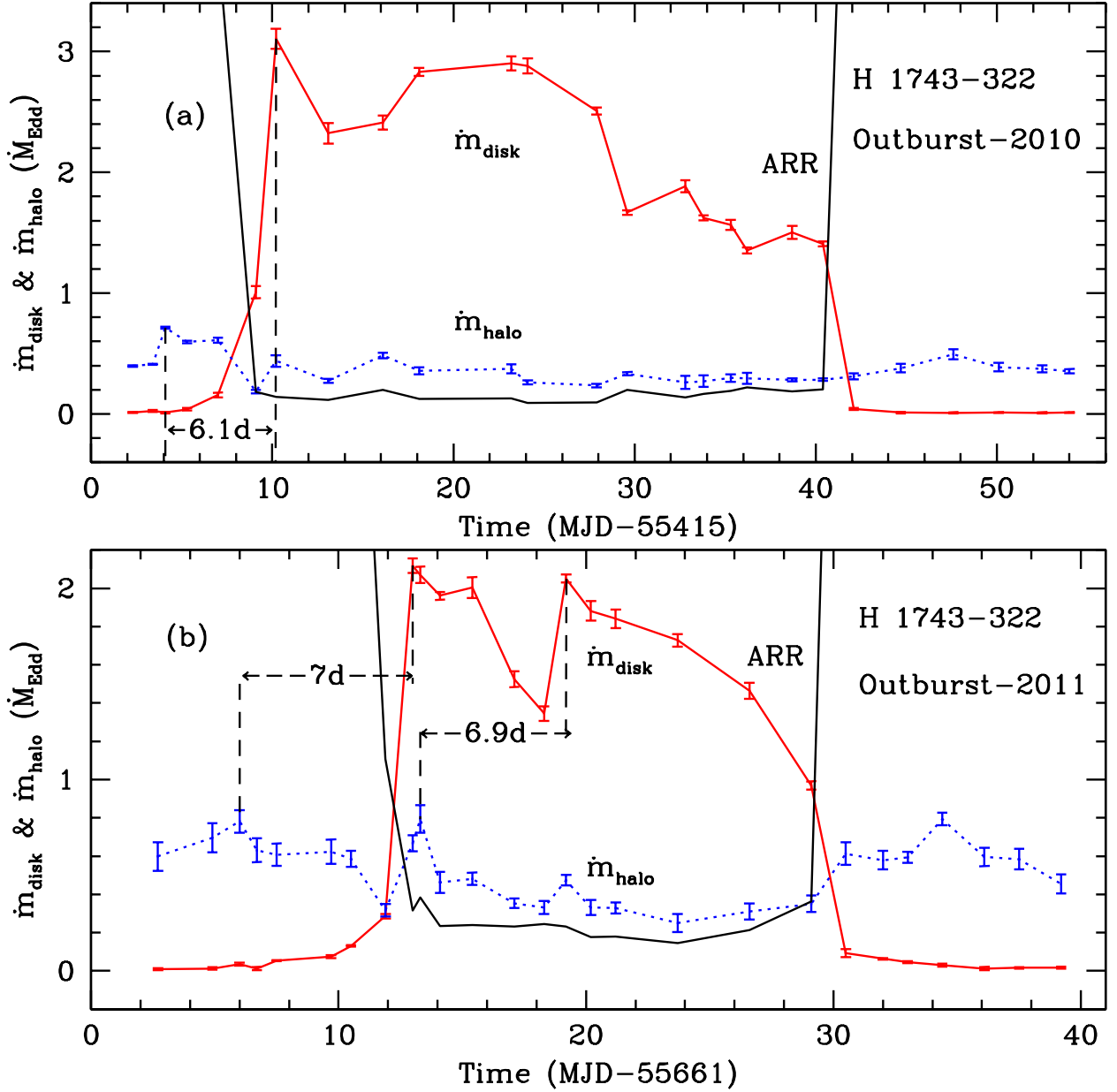}
\caption{Two outbursts in H 1743-322: Two accretion rates, $\dot m_{disc}$ and $\dot m_{halo}$, obtained using RXTE/PCA (2.5-25 keV) data by Molla, Chakrabarti, Debnath, \& Mondal (2017), are plotted (a) \& (b). Accretion rate ratio (ARR) is also shown in both cases, ignoring its higher values (with maxima at $\sim 60$) at two extreme points of the outbursts. The differences of peaking instants of the mass rates, or the time delays of about a week are marked in each box.}
\end{center}
\end{figure}


\begin{figure}
\begin{center}
\includegraphics[width=\columnwidth]{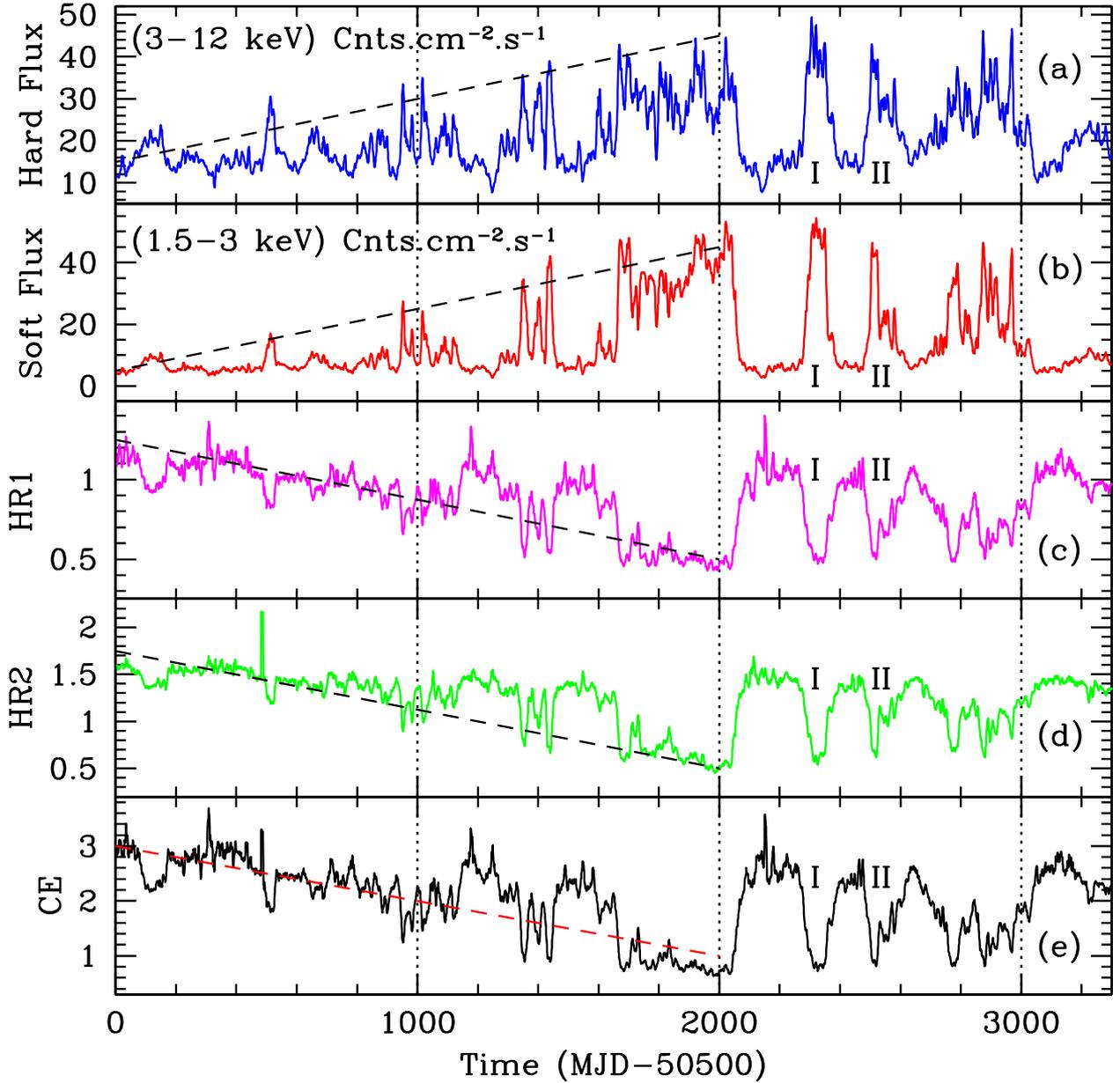}
\caption{Long-time behaviour of Cyg X-1: (a) hard flux (3-12 keV), (b) soft (1.5-3 keV) flux, (c) HR1 (=$b/a$), (d) HR2 (=$c/b$), and (e) CE [=($b+c$)/$a$]. Weekly running-mean data of RXTE/ASM are used for a quick look. The soft flux slowly becomes comparable to the hard flux, and HRs \& CE gradually decrease during the first $2000$ days. A softening spectrum towards hard-/soft- intermediate state (indicated by dashed straight lines) is apparent. Thereafter, the spectrum becomes hard for about $100$ days, and a uniquely prominent outburst-like flare (I), with similar photon fluxes, occurs. The `transience', with decreasing luminosity, still continues in an irregular manner with a weak flare (II) among others following it.}
\end{center}
\end{figure}


\begin{figure}
\begin{center}
\includegraphics[width=\columnwidth]{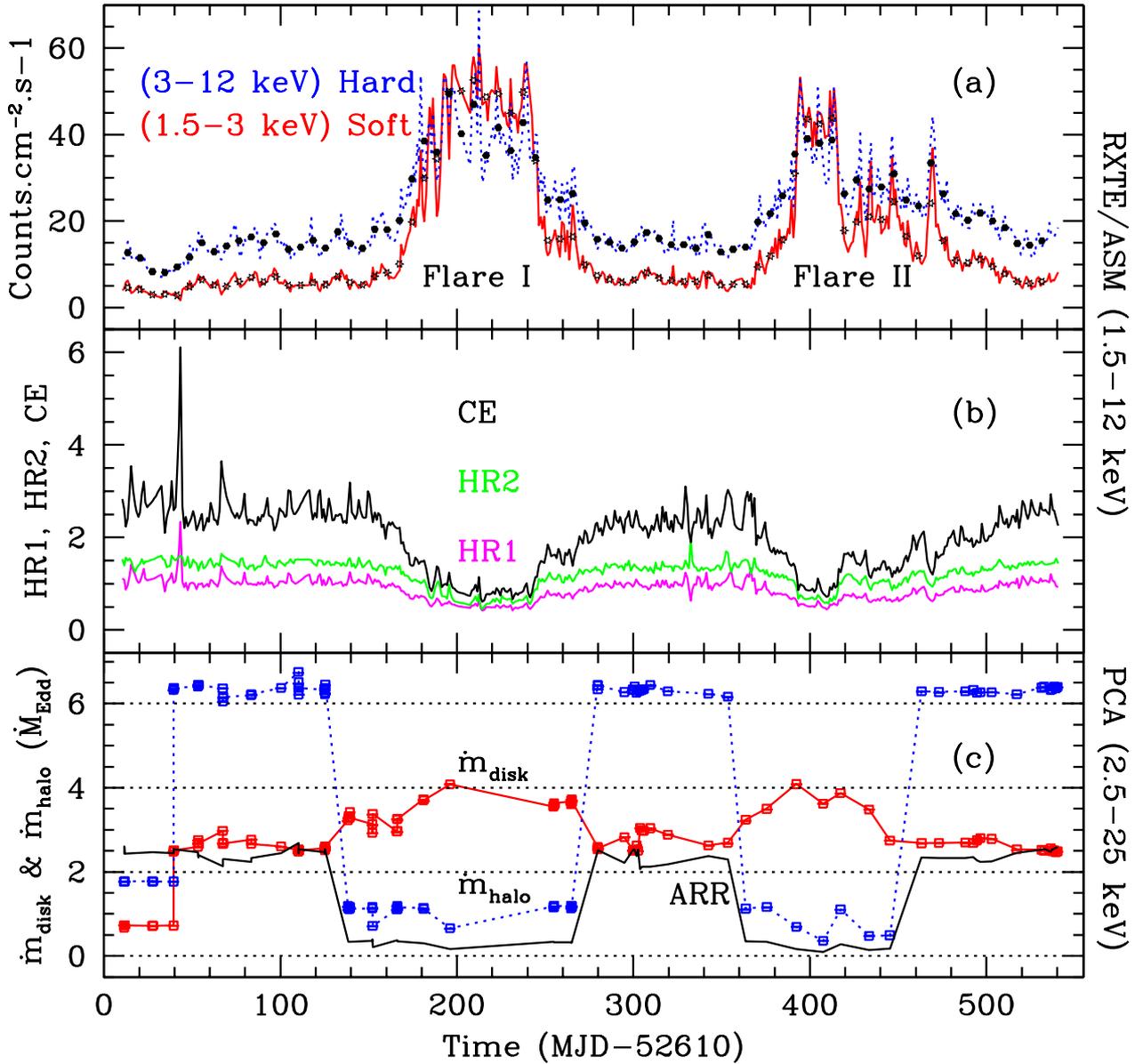}
\caption{Flares in Cyg X-1: (a) Long-time lightcurves (dotted/online-blue for hard and continuous/online-red for soft) with RXTE/ASM daily data are plotted.  However, weekly average data are plotted with points (dots for hard \& stars for soft) for a quick look at their remarkable synchronicity; both hard and soft fluxes seem to evolve simultaneously. Comptonizing efficiency (CE), and two hardness ratios defined here as HR1=(b/a) \& HR2=(c/b) (a, b, c being the three counts of the three bands in ASM) are plotted in (b). All quantities are seen to vary in a similar manner. In (c), two accretion rates $\dot m_{disc}$ and $\dot m_{halo}$, obtained from TCAF fits using RXTE/PCA (2.5-25 keV) data, are plotted. The accretion rate ratio (ARR) is practically constant during the flares. Both mass rates quite possibly peak at the same instant ($\sim$ MJD 52650) before the prominent Flare I indicating no apparent time delay ($0d^{+4.42}_{-1.43}h$), in agreement with the prediction of Ghosh \& Chakrabarti (2018). Flare II, occurred immediately after Flare I without any tangible achievement of hard state in between, is relatively less luminous and irregular.}
\end{center}
\end{figure}

\begin{figure}
\begin{center}
\includegraphics[width=\columnwidth]{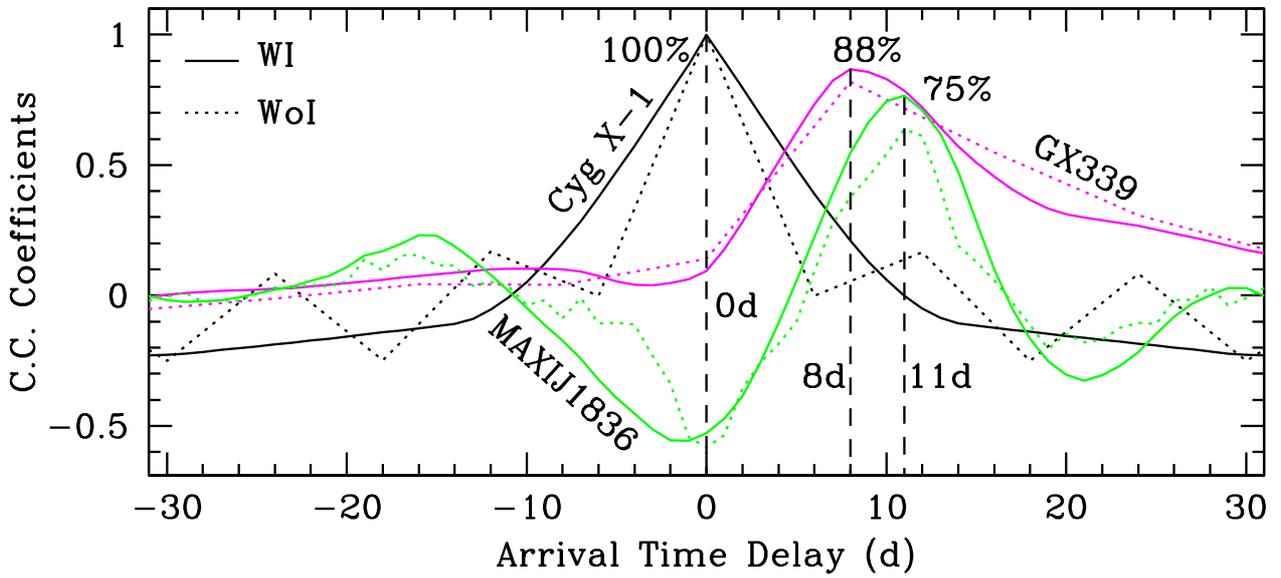}
\caption{Fine-tuned Flare-I of Cyg X-1 vs. outbursts: Two mass accretion rates are cross-correlated, where the time window spans between the commencement and the peak of the flare/outburst. Time lags are similar to those estimated earlier. Correlation coefficients of the relevant peaks are marked in percentage for comparison. (WI=with interpolation \& WoI=without interpolation)}
\end{center}
\end{figure}


  



\begin{landscape}
\begin{table}
\tiny
\vspace{-1.7cm}
\vskip0.05cm
\begin{flushleft}
\caption{Flare-I in Cyg X-1: Variation of the TCAF fit parameters, keeping all the parameters free, in the 2.5-25 keV
energy range.}
\begin{center}
\begin{tabular}{c c c c c c c c c c c}
\hline
\hline
&&&&&&&&&&\\
$Obs$ & $\rm Id$ & $\rm MJD$ & $ N_H $ & $\rm \dot{m}_{disk} ~ (\dot{M}_{Edd})$ & $\rm \dot{m}_{halo} ~(\dot{M}_{Edd})$ & $M_{BH} ~ (M_{\odot})$ & $X_s ~ (r_g)$ & $ R $ & $N$ & $\chi^2/dof$ \\\\
\hline 
$1$ & $60090-01-21-01$ & $52621.5094$ & $0.510$ & $0.732^{+0.004}_{-0.004}$ & $1.770^{+0.003} _{-0.003}$ & $14.048^{+0.034}_{-0.210}$ & $72.644^{+1.714}_{-0.555}$ & $1.205^{+0.001}_{-0.001}$ & $1.863^{+0.026}_{-0.026}$ & $41.76/39$ \\
$2$ & $60090-01-21-02$ & $52621.5859$ & $0.480$ & $0.678^{+0.014}_{-0.014}$ & $1.770^{+0.001} _{-0.001}$ & $13.962^{+0.029}_{-0.244}$ & $72.785^{+0.839}_{-0.913}$ & $1.205^{+0.001}_{-0.001}$ & $2.280^{+0.029}_{-0.029}$ & $56.68/39$ \\
$3$ & $60090-01-21-04$ & $52621.7170$ & $0.752$ & $0.730^{+0.004}_{-0.004}$ & $1.773^{+0.004} _{-0.004}$ & $14.012^{+0.036}_{-0.233}$ & $72.607^{+1.776}_{-0.676}$ & $1.204^{+0.001}_{-0.001}$ & $2.013^{+0.030}_{-0.030}$ & $52.37/39$ \\
$4$ & $60090-01-22-00$ & $52637.4654$ & $0.480$ & $0.719^{+0.003}_{-0.013}$ & $1.775^{+0.001} _{-0.002}$ & $14.029^{+0.026}_{-0.203}$ & $72.606^{+1.251}_{-0.568}$ & $1.205^{+0.001}_{-0.001}$ & $2.298^{+0.028}_{-0.028}$ & $57.23/39$ \\
$5$ & $70414-01-05-01$ & $52637.9015$ & $1.057$ & $0.716^{+0.005}_{-0.011}$ & $1.765^{+0.001} _{-0.001}$ & $14.497^{+0.061}_{-0.224}$ & $73.082^{+1.786}_{-0.541}$ & $1.203^{+0.001}_{-0.001}$ & $1.717^{+0.018}_{-0.018}$ & $64.80/39$ \\
$6$ & $60090-01-23-02$ & $52649.4254$ & $0.841$ & $0.725^{+0.003}_{-0.003}$ & $1.774^{+0.003} _{-0.003}$ & $14.028^{+0.027}_{-0.194}$ & $72.623^{+1.333}_{-0.573}$ & $1.204^{+0.001}_{-0.001}$ & $1.489^{+0.018}_{-0.018}$ & $52.84/39$ \\
$7$ & $60090-01-23-01$ & $52649.4852$ & $1.984$ & $2.483^{+0.012}_{-0.011}$ & $6.325^{+0.009} _{-0.009}$ & $15.201^{+0.305}_{-0.313}$ & $199.749^{+2.028}_{-2.108}$ & $2.968^{+0.001}_{-0.002}$ & $0.301^{+0.002}_{-0.002}$ & $33.83/39$ \\
$8$ & $60090-01-23-00$ & $52649.6694$ & $1.600$ & $2.512^{+0.010}_{-0.010}$ & $6.368^{+0.008} _{-0.008}$ & $14.838^{+0.190}_{-0.180}$ & $199.918^{+1.671}_{-1.555}$ & $2.955^{+0.001}_{-0.001}$ & $0.385^{+0.002}_{-0.002}$ & $27.45/39$ \\
$9$ & $60090-01-24-00$ & $52663.3831$ & $0.480$ & $2.596^{+0.013}_{-0.012}$ & $6.405^{+0.009} _{-0.009}$ & $13.455^{+0.111}_{-0.032}$ & $201.614^{+0.122}_{-1.733}$ & $2.939^{+0.001}_{-0.001}$ & $1.0453^{+0.007}_{-0.007}$ & $50.91/39$ \\
$10$ & $60090-01-24-02$ & $52663.4519$ & $1.242$ & $2.679^{+0.017}_{-0.015}$ & $6.429^{+0.011} _{-0.003}$ & $14.530^{+0.588}_{-0.404}$ & $186.553^{+2.027}_{-2.074}$ & $2.932^{+0.001}_{-0.001}$ & $1.026^{+0.009}_{-0.009}$ & $23.26/39$ \\
$11$ & $60090-01-24-01$ & $52663.5143$ & $1.074$ & $2.753^{+0.013}_{-0.013}$ & $6.440^{+0.002} _{-0.002}$ & $14.663^{+0.697}_{-0.400}$ & $186.857^{+2.012}_{-2.382}$ & $2.934^{+0.001}_{-0.001}$ & $1.158^{+0.011}_{-0.010}$ & $48.99/39$ \\
$12$ & $60090-01-24-03$ & $52663.6124$ & $1.256$ & $2.681^{+0.014}_{-0.006}$ & $6.445^{+0.026} _{-0.002}$ & $14.525^{+0.505}_{-0.373}$ & $186.334^{+1.848}_{-1.818}$ & $2.931^{+0.001}_{-0.001}$ & $1.106^{+0.009}_{-0.009}$ & $43.30/39$ \\
$13$ & $60090-01-25-00$ & $52677.3372$ & $1.490$ & $2.976^{+0.010}_{-0.010}$ & $6.362^{+0.002} _{-0.002}$ & $14.952^{+0.171}_{-0.144}$ & $194.314^{+1.102}_{-1.147}$ & $2.953^{+0.002}_{-0.002}$ & $0.599^{+0.004}_{-0.004}$ & $33.15/39$ \\
$14$ & $60090-01-25-01$ & $52677.3991$ & $0.802$ & $2.680^{+0.015}_{-0.014}$ & $6.047^{+0.006} _{-0.005}$ & $15.676^{+0.104}_{-0.256}$ & $200.343^{+1.439}_{-1.092}$ & $2.978^{+0.115}_{-0.023}$ & $0.351^{+0.124}_{-0.018}$ & $34.94/39$ \\
$15$ & $60090-01-25-02$ & $52677.4685$ & $1.566$ & $2.682^{+0.012}_{-0.016}$ & $6.145^{+0.009} _{-0.003}$ & $14.683^{+0.423}_{-0.325}$ & $193.357^{+2.378}_{-2.381}$ & $2.980^{+0.329}_{-0.011}$ & $0.482^{+0.358}_{-0.013}$ & $29.11/39$ \\
$16$ & $60090-01-25-04$ & $52677.6004$ & $1.231$ & $2.671^{+0.033}_{-0.018}$ & $6.154^{+0.010} _{-0.015}$ & $14.373^{+0.427}_{-0.319}$ & $193.923^{+2.753}_{-2.758}$ & $2.986^{+0.165}_{-0.017}$ & $0.571^{+0.307}_{-0.023}$ & $48.83/39$ \\
$17$ & $60090-01-26-00$ & $52693.5206$ & $0.867$ & $2.771^{+0.009}_{-0.009}$ & $6.202^{+0.002} _{-0.002}$ & $14.772^{+0.148}_{-0.152}$ & $200.545^{+1.201}_{-1.367}$ & $2.952^{+0.001}_{-0.001}$ & $0.574^{+0.003}_{-0.003}$ & $46.59/39$ \\
$18$ & $60090-01-26-01$ & $52693.5824$ & $1.801$ & $2.668^{+0.037}_{-0.013}$ & $6.220^{+0.007} _{-0.012}$ & $14.729^{+0.259}_{-0.223}$ & $194.858^{+1.702}_{-1.674}$ & $2.956^{+0.001}_{-0.001}$ & $0.672^{+0.004}_{-0.004}$ & $30.99/39$ \\
$19$ & $60090-01-27-01$ & $52710.3643$ & $1.438$ & $2.605^{+0.023}_{-0.022}$ & $6.372^{+0.014} _{-0.014}$ & $14.575^{+0.645}_{-0.447}$ & $195.257^{+3.948}_{-3.884}$ & $2.940^{+0.003}_{-0.003}$ & $0.698^{+0.008}_{-0.008}$ & $53.95/39$ \\
$20$ & $60090-01-28-00$ & $52720.2387$ & $1.231$ & $2.511^{+0.004}_{-0.004}$ & $6.758^{+0.024} _{-0.024}$ & $14.825^{+0.213}_{-0.192}$ & $199.753^{+1.817}_{-1.748}$ & $2.945^{+0.002}_{-0.002}$ & $0.611^{+0.004}_{-0.004}$ & $41.50/39$ \\
$21$ & $60090-01-28-02$ & $52720.2948$ & $1.459$ & $2.482^{+0.004}_{-0.004}$ & $6.507^{+0.031} _{-0.031}$ & $14.571^{+0.308}_{-0.263}$ & $195.873^{+2.285}_{-2.181}$ & $2.952^{+0.002}_{-0.002}$ & $0.689^{+0.006}_{-0.006}$ & $39.87/39$ \\
$22$ & $60090-01-28-01$ & $52720.3713$ & $1.020$ & $2.539^{+0.014}_{-0.014}$ & $6.219^{+0.008} _{-0.008}$ & $15.058^{+0.356}_{-0.304}$ & $198.032^{+2.466}_{-2.353}$ & $2.971^{+0.022}_{-0.003}$ & $0.484^{+0.018}_{-0.004}$ & $27.96/39$ \\
$23$ & $60090-01-28-03$ & $52720.4365$ & $1.388$ & $2.510^{+0.014}_{-0.014}$ & $6.373^{+0.010} _{-0.010}$ & $14.627^{+0.337}_{-0.283}$ & $197.628^{+2.608}_{-2.495}$ & $2.954^{+0.002}_{-0.002}$ & $0.558^{+0.004}_{-0.004}$ & $29.39/39$ \\
$24$ & $60090-01-29-00$ & $52735.1615$ & $1.218$ & $2.563^{+0.013}_{-0.012}$ & $6.328^{+0.009} _{-0.009}$ & $14.593^{+0.249}_{-0.217}$ & $195.290^{+1.797}_{-1.739}$ & $2.953^{+0.001}_{-0.001}$ & $0.728^{+0.005}_{-0.005}$ & $37.09/39$ \\
$25$ & $60090-01-29-03$ & $52735.2241$ & $1.264$ & $2.515^{+0.013}_{-0.013}$ & $6.375^{+0.010} _{-0.010}$ & $14.622^{+0.291}_{-0.250}$ & $196.419^{+2.167}_{-2.083}$ & $2.955^{+0.002}_{-0.002}$ & $0.672^{+0.005}_{-0.005}$ & $18.46/39$ \\
$26$ & $60090-01-29-04$ & $52735.2948$ & $1.028$ & $2.557^{+0.014}_{-0.013}$ & $6.229^{+0.009} _{-0.009}$ & $14.995^{+0.355}_{-0.301}$ & $195.239^{+2.191}_{-2.098}$ & $2.971^{+0.020}_{-0.002}$ & $0.609^{+0.021}_{-0.004}$ & $39.39/39$ \\
$27$ & $60090-01-29-01$ & $52735.3706$ & $1.144$ & $2.532^{+0.013}_{-0.013}$ & $6.257^{+0.008} _{-0.008}$ & $15.040^{+0.336}_{-0.289}$ & $198.057^{+2.352}_{-2.244}$ & $2.969^{+0.038}_{-0.002}$ & $0.547^{+0.038}_{-0.003}$ & $36.52/39$ \\
$28$ & $60090-01-29-02$ & $52735.4317$ & $1.389$ & $2.602^{+0.008}_{-0.004}$ & $6.458^{+0.024} _{-0.014}$ & $14.700^{+0.198}_{-0.178}$ & $198.261^{+1.639}_{-1.592}$ & $2.918^{+0.001}_{-0.001}$ & $0.577^{+0.003}_{-0.003}$ & $23.29/39$ \\
$29$ & $60090-01-30-00$ & $52748.2531$ & $2.565$ & $3.225^{+0.077}_{-0.072}$ & $1.185^{+0.019} _{-0.020}$ & $14.649^{+0.038}_{-0.038}$ & $34.660^{+0.308}_{-0.308}$ & $1.097^{+0.001}_{-0.001}$ & $1.053^{+0.012}_{-0.012}$ & $26.13/39$ \\
$30$ & $80111-01-01-02$ & $52748.5035$ & $2.360$ & $3.289^{+0.070}_{-0.034}$ & $1.120^{+0.016} _{-0.030}$ & $14.500^{+0.078}_{-0.077}$ & $40.277^{+0.305}_{-0.307}$ & $1.095^{+0.001}_{-0.001}$ & $1.545^{+0.020}_{-0.020}$ & $18.52/39$ \\
$31$ & $80111-01-01-00$ & $52749.3067$ & $4.522$ & $3.419^{+0.064}_{-0.061}$ & $1.167^{+0.011} _{-0.009}$ & $16.129^{+0.043}_{-0.043}$ & $44.127^{+0.228}_{-0.229}$ & $1.080^{+0.001}_{-0.001}$ & $1.743^{+0.020}_{-0.020}$ & $25.10/39$ \\
$32$ & $80111-01-01-03$ & $52749.3726$ & $3.103$ & $3.292^{+0.053}_{-0.051}$ & $1.126^{+0.007} _{-0.007}$ & $17.011^{+0.073}_{-0.072}$ & $43.788^{+0.243}_{-0.245}$ & $1.081^{+0.001}_{-0.001}$ & $1.842^{+0.024}_{-0.024}$ & $32.52/39$ \\
$33$ & $60090-01-31-00$ & $52762.1331$ & $1.845$ & $3.135^{+0.062}_{-0.058}$ & $1.137^{+0.016} _{-0.016}$ & $14.500^{+0.059}_{-0.059}$ & $36.747^{+0.324}_{-0.325}$ & $1.093^{+0.001}_{-0.001}$ & $1.146^{+0.014}_{-0.014}$ & $23.89/39$ \\
$34$ & $60090-01-31-01$ & $52762.1983$ & $2.038$ & $2.934^{+0.064}_{-0.060}$ & $1.161^{+0.017} _{-0.017}$ & $14.536^{+0.086}_{-0.087}$ & $34.464^{+0.513}_{-0.515}$ & $1.083^{+0.001}_{-0.001}$ & $1.141^{+0.021}_{-0.020}$ & $31.54/39$ \\
$35$ & $60090-01-31-02$ & $52762.3007$ & $2.455$ & $3.378^{+0.013}_{-0.026}$ & $0.721^{+0.012} _{-0.017}$ & $14.500^{+0.067}_{-0.066}$ & $49.203^{+0.174}_{-0.175}$ & $1.074^{+0.001}_{-0.001}$ & $4.465^{+0.097}_{-0.096}$ & $26.09/39$ \\
$36$ & $60090-01-32-00$ & $52776.0804$ & $1.654$ & $2.969^{+0.043}_{-0.026}$ & $1.113^{+0.012} _{-0.026}$ & $15.741^{+0.094}_{-0.113}$ & $35.116^{+0.352}_{-0.358}$ & $1.082^{+0.001}_{-0.001}$ & $1.332^{+0.017}_{-0.017}$ & $23.50/39$ \\
$37$ & $60090-01-32-01$ & $52776.1456$ & $1.523$ & $2.966^{+0.045}_{-0.026}$ & $1.115^{+0.013} _{-0.025}$ & $15.408^{+0.114}_{-0.111}$ & $33.892^{+0.368}_{-0.363}$ & $1.083^{+0.001}_{-0.001}$ & $1.513^{+0.020}_{-0.019}$ & $26.52/39$ \\
$38$ & $60090-01-32-02$ & $52776.2122$ & $1.500$ & $3.267^{+0.065}_{-0.058}$ & $1.193^{+0.016} _{-0.016}$ & $14.048^{+0.062}_{-0.061}$ & $30.949^{+0.348}_{-0.334}$ & $1.091^{+0.001}_{-0.001}$ & $1.583^{+0.017}_{-0.017}$ & $16.12/39$ \\
$39$ & $60090-01-32-03$ & $52776.2776$ & $1.294$ & $3.249^{+0.069}_{-0.062}$ & $1.151^{+0.018} _{-0.019}$ & $14.369^{+0.084}_{-0.082}$ & $31.782^{+0.373}_{-0.361}$ & $1.094^{+0.001}_{-0.001}$ & $1.776^{+0.021}_{-0.020}$ & $21.59/39$ \\
$40$ & $60090-01-31-03$ & $52791.0209$ & $0.563$ & $3.724^{+0.084}_{-0.064}$ & $1.142^{+0.016} _{-0.013}$ & $14.635^{+0.111}_{-0.304}$ & $29.558^{+0.310}_{-0.226}$ & $1.103^{+0.002}_{-0.001}$ & $1.972^{+0.014}_{-0.015}$ & $43.11/39$ \\
$41$ & $60090-01-33-00$ & $52791.2713$ & $0.770$ & $3.704^{+0.062}_{-0.033}$ & $1.121^{+0.013} _{-0.021}$ & $15.281^{+0.096}_{-0.226}$ & $29.545^{+0.272}_{-0.173}$ & $1.115^{+0.001}_{-0.001}$ & $1.776^{+0.013}_{-0.013}$ & $52.56/39$ \\
$42$ & $60090-01-34-03$ & $52806.0594$ & $1.964$ & $4.084^{+0.002}_{-0.002}$ & $0.659^{+0.004} _{-0.004}$ & $14.501^{+0.044}_{-0.077}$ & $29.651^{+0.154}_{-0.144}$ & $1.224^{+0.003}_{-0.003}$ & $17.555^{+0.083}_{-0.078}$ & $76.77/39$ \\
$43$ & $60090-01-38-00$ & $52864.8337$ & $1.007$ & $3.550^{+0.053}_{-0.069}$ & $1.182^{+0.010} _{-0.014}$ & $13.884^{+0.240}_{-0.352}$ & $29.192^{+0.214}_{-0.323}$ & $1.094^{+0.001}_{-0.001}$ & $1.624^{+0.013}_{-0.014}$ & $27.24/39$ \\
$44$ & $60090-01-38-03$ & $52864.9354$ & $0.730$ & $3.578^{+0.057}_{-0.063}$ & $1.153^{+0.012} _{-0.014}$ & $15.095^{+0.104}_{-0.345}$ & $29.635^{+0.292}_{-0.327}$ & $1.092^{+0.001}_{-0.001}$ & $2.065^{+0.018}_{-0.017}$ & $33.89/39$ \\
$45$ & $60090-01-38-01$ & $52865.0028$ & $0.683$ & $3.595^{+0.058}_{-0.057}$ & $1.189^{+0.010} _{-0.010}$ & $14.341^{+0.228}_{-0.412}$ & $29.401^{+0.335}_{-0.316}$ & $1.089^{+0.001}_{-0.001}$ & $1.980^{+0.017}_{-0.016}$ & $40.06/39$ \\
$46$ & $60090-01-38-02$ & $52865.0702$ & $0.570$ & $3.634^{+0.053}_{-0.054}$ & $1.185^{+0.009} _{-0.010}$ & $14.249^{+0.476}_{-0.561}$ & $28.924^{+0.319}_{-0.395}$ & $1.087^{+0.001}_{-0.001}$ & $2.469^{+0.020}_{-0.018}$ & $45.26/39$ \\
$47$ & $60090-01-39-00$ & $52874.7678$ & $0.569$ & $3.680^{+0.055}_{-0.051}$ & $1.188^{+0.009} _{-0.009}$ & $14.525^{+0.461}_{-0.548}$ & $29.156^{+0.287}_{-0.358}$ & $1.086^{+0.001}_{-0.001}$ & $2.177^{+0.017}_{-0.016}$ & $39.83/39$ \\
$48$ & $60090-01-39-04$ & $52874.8317$ & $0.493$ & $3.609^{+0.058}_{-0.050}$ & $1.131^{+0.012} _{-0.011}$ & $15.557^{+0.120}_{-0.350}$ & $29.622^{+0.282}_{-0.280}$ & $1.091^{+0.001}_{-0.001}$ & $2.490^{+0.020}_{-0.018}$ & $55.89/39$ \\
$49$ & $60090-01-39-03$ & $52874.8991$ & $0.522$ & $3.647^{+0.065}_{-0.049}$ & $1.194^{+0.011} _{-0.008}$ & $14.453^{+0.145}_{-0.384}$ & $29.489^{+0.342}_{-0.270}$ & $1.088^{+0.001}_{-0.001}$ & $1.761^{+0.016}_{-0.013}$ & $28.30/39$ \\
$50$ & $60090-01-39-02$ & $52874.9657$ & $0.718$ & $3.719^{+0.056}_{-0.052}$ & $1.152^{+0.010} _{-0.010}$ & $15.868^{+0.104}_{-0.511}$ & $29.551^{+0.301}_{-0.292}$ & $1.088^{+0.001}_{-0.001}$ & $2.749^{+0.021}_{-0.020}$ & $42.27/39$ \\
$51$ & $60090-01-39-01$ & $52875.0352$ & $0.618$ & $3.695^{+0.055}_{-0.054}$ & $1.174^{+0.009} _{-0.009}$ & $15.074^{+0.442}_{-0.684}$ & $29.342^{+0.429}_{-0.388}$ & $1.085^{+0.001}_{-0.001}$ & $2.449^{+0.020}_{-0.020}$ & $35.84/39$ \\
$52$ & $60090-01-40-00$ & $52889.7476$ & $1.140$ & $2.593^{+0.012}_{-0.009}$ & $6.444^{+0.020} _{-0.009}$ & $14.569^{+0.268}_{-0.223}$ & $195.363^{+1.836}_{-1.846}$ & $2.934^{+0.002}_{-0.001}$ & $0.567^{+0.005}_{-0.004}$ & $36.62/39$ \\
$53$ & $60090-01-40-01$ & $52889.8165$ & $0.860$ & $2.537^{+0.414}_{-0.013}$ & $6.345^{+0.009} _{-0.009}$ & $14.853^{+0.338}_{-0.274}$ & $197.256^{+2.211}_{-2.235}$ & $2.967^{+0.001}_{-0.001}$ & $0.579^{+0.004}_{-0.004}$ & $41.52/39$ \\
$54$ & $60090-01-40-03$ & $52890.0137$ & $1.179$ & $2.560^{+0.011}_{-0.010}$ & $6.434^{+0.009} _{-0.009}$ & $14.465^{+0.238}_{-0.187}$ & $193.817^{+1.480}_{-1.552}$ & $2.950^{+0.001}_{-0.001}$ & $0.739^{+0.005}_{-0.005}$ & $47.16/39$ \\
$55$ & $60090-01-41-00$ & $52904.9778$ & $0.781$ & $2.826^{+0.009}_{-0.009}$ & $6.275^{+0.003} _{-0.002}$ & $14.826^{+0.127}_{-0.143}$ & $200.649^{+1.079}_{-1.195}$ & $2.946^{+0.001}_{-0.001}$ & $0.518^{+0.003}_{-0.003}$ & $39.57/39$ \\
\hline
\end{tabular}
\end{center}
\end{flushleft}
\label{tcaf_fits:nfree}

\end{table}
\end{landscape}



\begin{landscape}
\begin{table}
\tiny
\vspace{-0.5cm}
\vskip0.05cm
\begin{flushleft}
\caption{Flare-II in Cyg X-1: Variation of the TCAF fit parameters, keeping all the parameters free, in the 2.5-25 keV
energy range.} 
\begin{center}
\begin{tabular}{c c c c c c c c c c c}
\hline
\hline
&&&&&&&&&&\\
$Obs$ & $\rm Id$ & $\rm MJD$ & $ N_H $ & $\rm \dot{m}_{disk} ~ (\dot{M}_{Edd})$ & $\rm \dot{m}_{halo} ~(\dot{M}_{Edd})$ & $M_{BH} ~ (M_{\odot})$ & $X_s ~ (r_g)$ & $ R $ & $N$ & $\chi^2/dof$ \\\\
\hline 
$1$ & $60090-01-53-00$ & $52909.751$ & $0.968$ & $2.508^{+0.011}_{-0.011}$ & $6.334^{+0.009} _{-0.009}$ & $14.741^{+0.223}_{-0.199}$ & $196.306^{+1.630}_{-1.573}$ & $2.967^{+0.001}_{-0.001}$ & $0.667^{+0.004}_{-0.004}$ & $33.75/39$ \\
$2$ & $60090-01-53-01$ & $52910.738$ & $1.219$ & $2.530^{+0.011}_{-0.011}$ & $6.402^{+0.009} _{-0.009}$ & $14.580^{+0.213}_{-0.189}$ & $195.698^{+1.575}_{-1.533}$ & $2.953^{+0.001}_{-0.001}$ & $0.665^{+0.004}_{-0.004}$ & $30.60/39$ \\
$3$ & $60090-01-53-02$ & $52911.723$ & $2.918$ & $2.633^{+0.012}_{-0.012}$ & $6.257^{+0.008} _{-0.008}$ & $14.962^{+0.417}_{-0.332}$ & $188.456^{+1.593}_{-1.582}$ & $2.968^{+0.001}_{-0.001}$ & $0.887^{+0.006}_{-0.006}$ & $51.29/39$ \\
$4$ & $60090-01-53-03$ & $52912.707$ & $1.337$ & $2.502^{+0.011}_{-0.011}$ & $6.317^{+0.008} _{-0.008}$ & $14.932^{+0.227}_{-0.206}$ & $198.345^{+1.704}_{-1.667}$ & $2.965^{+0.001}_{-0.002}$ & $0.537^{+0.003}_{-0.003}$ & $31.80/39$ \\
$5$ & $60090-01-53-04$ & $52913.760$ & $0.704$ & $3.043^{+0.009}_{-0.009}$ & $6.301^{+0.003} _{-0.003}$ & $14.745^{+0.101}_{-0.093}$ & $201.170^{+0.603}_{-1.115}$ & $3.018^{+0.027}_{-0.029}$ & $0.908^{+0.053}_{-0.053}$ & $49.20/39$ \\
$6$ & $60090-01-53-05$ & $52914.746$ & $0.780$ & $2.985^{+0.010}_{-0.010}$ & $6.327^{+0.002} _{-0.002}$ & $14.444^{+0.116}_{-0.100}$ & $194.138^{+0.933}_{-0.964}$ & $2.974^{+0.026}_{-0.005}$ & $1.118^{+0.057}_{-0.016}$ & $21.16/39$ \\
$7$ & $60090-01-42-01$ & $52915.797$ & $0.700$ & $2.979^{+0.010}_{-0.010}$ & $6.341^{+0.003} _{-0.003}$ & $14.887^{+0.179}_{-0.148}$ & $194.400^{+1.144}_{-1.204}$ & $2.965^{+0.002}_{-0.002}$ & $0.896^{+0.006}_{-0.006}$ & $32.18/39$ \\
$8$ & $60090-01-42-02$ & $52919.631$ & $0.629$ & $3.038^{+0.009}_{-0.009}$ & $6.447^{+0.036} _{-0.003}$ & $14.550^{+0.078}_{-0.071}$ & $200.063^{+0.899}_{-0.916}$ & $2.941^{+0.002}_{-0.002}$ & $0.704^{+0.005}_{-0.005}$ & $39.46/39$ \\
$9$ & $60090-01-43-00$ & $52929.620$ & $0.836$ & $2.887^{+0.010}_{-0.010}$ & $6.291^{+0.003} _{-0.003}$ & $14.692^{+0.145}_{-0.126}$ & $199.640^{+1.296}_{-1.326}$ & $2.964^{+0.002}_{-0.002}$ & $0.592^{+0.004}_{-0.004}$ & $32.78/39$ \\
$10$ & $60090-01-44-00$ & $52952.533$ & $0.574$ & $2.629^{+0.012}_{-0.011}$ & $6.236^{+0.008} _{-0.008}$ & $14.778^{+0.258}_{-0.219}$ & $194.639^{+1.588}_{-1.583}$ & $2.980^{+0.028}_{-0.007}$ & $0.685^{+0.039}_{-0.013}$ & $43.61/39$ \\
$11$ & $60090-01-45-00$ & $52963.500$ & $0.756$ & $2.694^{+0.038}_{-0.014}$ & $6.170^{+0.017} _{-0.017}$ & $14.377^{+0.149}_{-0.133}$ & $196.622^{+1.272}_{-1.264}$ & $3.113^{+0.057}_{-0.082}$ & $0.721^{+0.137}_{-0.139}$ & $33.29/39$ \\
$12$ & $60090-01-46-01$ & $52973.649$ & $2.483$ & $3.239^{+0.080}_{-0.056}$ & $1.128^{+0.021} _{-0.029}$ & $14.500^{+0.101}_{-0.101}$ & $35.2552^{+0.405}_{-0.406}$ & $1.097^{+0.001}_{-0.001}$ & $1.325^{+0.019}_{-0.019}$ & $35.11/39$ \\
$13$ & $60090-01-47-00$ & $52985.528$ & $1.201$ & $3.501^{+0.066}_{-0.064}$ & $1.166^{+0.014} _{-0.015}$ & $14.927^{+0.074}_{-0.074}$ & $31.6598^{+0.289}_{-0.291}$ & $1.094^{+0.001}_{-0.001}$ & $2.067^{+0.019}_{-0.019}$ & $30.98/39$ \\
$14$ & $60090-01-35-01$ & $53002.391$ & $0.500$ & $4.088^{+0.002}_{-0.002}$ & $0.695^{+0.004} _{-0.004}$ & $15.223^{+0.476}_{-0.395}$ & $29.3901^{+0.090}_{-0.089}$ & $1.217^{+0.003}_{-0.001}$ & $10.55^{+0.032}_{-0.032}$ & $50.92/39$ \\
$15$ & $60090-01-49-01$ & $53017.490$ & $0.480$ & $3.619^{+0.015}_{-0.015}$ & $0.364^{+0.002} _{-0.002}$ & $14.502^{+0.020}_{-0.019}$ & $37.8297^{+0.110}_{-0.111}$ & $1.086^{+0.001}_{-0.001}$ & $17.85^{+0.113}_{-0.112}$ & $42.55/39$ \\
$16$ & $60090-01-50-00$ & $53027.366$ & $0.480$ & $3.876^{+0.021}_{-0.004}$ & $1.109^{+0.010} _{-0.010}$ & $14.510^{+0.133}_{-0.132}$ & $42.6763^{+0.198}_{-0.200}$ & $1.088^{+0.001}_{-0.001}$ & $5.506^{+0.039}_{-0.039}$ & $60.06/39$ \\
$17$ & $60090-01-51-04$ & $53043.572$ & $1.199$ & $3.484^{+0.036}_{-0.039}$ & $0.479^{+0.001} _{-0.001}$ & $14.501^{+0.136}_{-0.126}$ & $30.3769^{+0.095}_{-0.090}$ & $1.165^{+0.001}_{-0.001}$ & $12.69^{+0.058}_{-0.057}$ & $47.93/39$ \\
$18$ & $60090-01-52-00$ & $53055.318$ & $0.772$ & $2.745^{+0.032}_{-0.032}$ & $0.493^{+0.001} _{-0.001}$ & $14.499^{+0.037}_{-0.036}$ & $29.4407^{+0.069}_{-0.069}$ & $1.100^{+0.001}_{-0.001}$ & $11.45^{+0.085}_{-0.084}$ & $53.80/39$ \\
$19$ & $80110-01-01-01$ & $53073.268$ & $1.375$ & $2.683^{+0.010}_{-0.014}$ & $6.290^{+0.010} _{-0.001}$ & $14.433^{+0.622}_{-0.661}$ & $184.920^{+2.874}_{-1.692}$ & $2.972^{+0.071}_{-0.002}$ & $1.486^{+0.263}_{-0.015}$ & $34.12/39$ \\
$20$ & $80110-01-02-01$ & $53083.176$ & $0.498$ & $2.689^{+0.010}_{-0.029}$ & $6.276^{+0.015} _{-0.001}$ & $14.714^{+0.765}_{-0.512}$ & $188.108^{+2.546}_{-2.513}$ & $2.977^{+0.087}_{-0.005}$ & $1.143^{+0.260}_{-0.017}$ & $43.48/39$ \\
$21$ & $60090-01-54-02$ & $53098.135$ & $0.480$ & $2.696^{+0.010}_{-0.047}$ & $6.291^{+0.021} _{-0.001}$ & $14.689^{+0.787}_{-0.516}$ & $187.795^{+2.628}_{-2.588}$ & $2.985^{+0.081}_{-0.015}$ & $1.348^{+0.287}_{-0.049}$ & $29.78/39$ \\
$22$ & $90126-01-01-01$ & $53102.598$ & $0.480$ & $2.692^{+0.009}_{-0.033}$ & $6.322^{+0.016} _{-0.010}$ & $14.356^{+0.475}_{-0.558}$ & $184.678^{+2.294}_{-1.314}$ & $2.978^{+0.056}_{-0.011}$ & $1.953^{+0.271}_{-0.057}$ & $42.42/39$ \\
$23$ & $90126-01-02-01$ & $53104.705$ & $0.529$ & $2.752^{+0.010}_{-0.010}$ & $6.259^{+0.002} _{-0.002}$ & $14.862^{+0.512}_{-0.327}$ & $189.730^{+1.761}_{-2.076}$ & $2.978^{+0.061}_{-0.005}$ & $1.044^{+0.152}_{-0.017}$ & $19.39/39$ \\
$24$ & $90126-01-03-00$ & $53106.598$ & $0.572$ & $2.803^{+0.010}_{-0.010}$ & $6.270^{+0.002} _{-0.002}$ & $14.623^{+0.254}_{-0.196}$ & $191.229^{+1.278}_{-1.389}$ & $2.979^{+0.043}_{-0.009}$ & $1.062^{+0.103}_{-0.026}$ & $32.05/39$ \\
$25$ & $80110-01-03-00$ & $53113.091$ & $0.937$ & $2.786^{+0.009}_{-0.009}$ & $6.268^{+0.001} _{-0.001}$ & $14.560^{+0.246}_{-0.190}$ & $190.713^{+1.224}_{-1.327}$ & $2.977^{+0.043}_{-0.007}$ & $1.053^{+0.103}_{-0.021}$ & $31.94/39$ \\
$26$ & $80110-01-04-02$ & $53127.234$ & $0.857$ & $2.535^{+0.015}_{-0.015}$ & $6.228^{+0.014} _{-0.014}$ & $15.333^{+0.311}_{-0.287}$ & $199.955^{+1.794}_{-2.153}$ & $3.052^{+0.017}_{-0.018}$ & $0.637^{+0.022}_{-0.022}$ & $30.55/39$ \\
$27$ & $80110-01-05-02$ & $53141.207$ & $2.040$ & $2.530^{+0.013}_{-0.013}$ & $6.380^{+0.009} _{-0.009}$ & $14.765^{+0.249}_{-0.221}$ & $197.957^{+1.935}_{-1.891}$ & $2.941^{+0.002}_{-0.002}$ & $0.624^{+0.005}_{-0.005}$ & $33.70/39$ \\
$28$ & $80113-01-01-00$ & $53142.503$ & $1.279$ & $2.508^{+0.011}_{-0.010}$ & $6.403^{+0.008} _{-0.008}$ & $14.776^{+0.150}_{-0.144}$ & $199.827^{+1.399}_{-1.318}$ & $2.937^{+0.001}_{-0.001}$ & $0.677^{+0.004}_{-0.004}$ & $27.55/39$ \\
$29$ & $80113-01-02-01$ & $53146.505$ & $2.570$ & $2.556^{+0.012}_{-0.012}$ & $6.334^{+0.009} _{-0.009}$ & $14.771^{+0.203}_{-0.184}$ & $195.420^{+1.454}_{-1.406}$ & $2.946^{+0.001}_{-0.001}$ & $0.760^{+0.005}_{-0.005}$ & $34.49/39$ \\
$30$ & $90414-03-01-00$ & $53147.327$ & $1.170$ & $2.497^{+0.014}_{-0.014}$ & $6.344^{+0.010} _{-0.010}$ & $14.837^{+0.311}_{-0.270}$ & $199.078^{+2.506}_{-2.372}$ & $2.949^{+0.002}_{-0.002}$ & $0.787^{+0.006}_{-0.006}$ & $54.68/39$ \\
$31$ & $90414-03-01-01$ & $53148.372$ & $1.446$ & $2.512^{+0.013}_{-0.012}$ & $6.405^{+0.009} _{-0.009}$ & $14.801^{+0.254}_{-0.350}$ & $199.648^{+2.112}_{-2.037}$ & $2.940^{+0.002}_{-0.002}$ & $0.663^{+0.005}_{-0.005}$ & $37.44/39$ \\
$32$ & $90414-03-01-02$ & $53149.364$ & $1.675$ & $2.466^{+0.026}_{-0.026}$ & $6.367^{+0.019} _{-0.019}$ & $14.893^{+0.884}_{-0.232}$ & $200.619^{+1.207}_{-5.684}$ & $2.964^{+0.005}_{-0.005}$ & $0.675^{+0.010}_{-0.010}$ & $32.39/39$ \\
$33$ & $90414-03-01-03$ & $53150.344$ & $0.969$ & $2.487^{+0.014}_{-0.014}$ & $6.395^{+0.010} _{-0.010}$ & $14.742^{+0.268}_{-0.237}$ & $199.322^{+2.292}_{-2.177}$ & $2.941^{+0.002}_{-0.002}$ & $0.725^{+0.005}_{-0.005}$ & $31.69/39$ \\

\hline
\end{tabular}
\end{center}
\end{flushleft}
\label{tcaf_fits:n1free}
\end{table}
\end{landscape}

\end{document}